\documentclass[a4paper,11pt]{article}
\usepackage{pos}

\title{Inhomogeneous phases in the 3+1-dimensional Nambu-Jona-Lasinio model and their dependence on the regularization scheme}
\ShortTitle{Inhomogeneous phases in the 3+1-dimensional Nambu-Jona-Lasinio model}

\author*[a]{Laurin Pannullo}

\author[a,b]{Marc Wagner}

\author[a]{Marc Winstel}

\affiliation[a]{Institut für Theoretische Physik, Goethe Universit\"at Frankfurt am Main \\
	Max-von-Laue-Str. 1, 60438 Frankfurt am Main, Germany}
\affiliation[b]{Helmholtz Research Academy Hesse for FAIR,\\
  Campus Riedberg, Max-von-Laue-Stra{\ss}e 12, D-60438 Frankfurt am Main, Germany}

\emailAdd{pannullo@itp.uni-frankfurt.de}
\emailAdd{mwagner@itp.uni-frankfurt.de}
\emailAdd{winstel@itp.uni-frankfurt.de}

\abstract{
In this work we study the $3+1$-dimensional Nambu-Jona-Lasinio (NJL) model in the mean field-approximation. We carry out calculations using five different regularization schemes (two continuum and three lattice regularization schemes) with particular focus on inhomogeneous phases and condensates. 
The regularization schemes lead to drastically different inhomogeneous regions.
We provide evidence that inhomogeneous condensates appear for all regularization schemes almost exclusively at values of the chemical potential and with wave numbers, which are of the order of or even larger than the corresponding regulators.
This can be interpreted as indication that inhomogeneous phases in the $3+1$-dimensional NJL model are rather artifacts of the regularization and not a consequence of the NJL Lagrangian and its symmetries. 
}

\FullConference{%
	The 39th International Symposium on Lattice Field Theory,\\
	8th-13th August, 2022,\\
	Rheinische Friedrich-Wilhelms-Universität Bonn, Bonn, Germany
}

\usepackage{hyperref}
\usepackage[capitalise]{cleveref}
\usepackage{slashed}
\usepackage{graphicx}
\usepackage{subcaption}
\usepackage{upgreek}
\usepackage{mathtools}
\usepackage{amsmath}
\usepackage{xcolor}
\usepackage{dsfont}
\usepackage{bbold}
\usepackage[utf8]{inputenc}
\usepackage[ngerman, english]{babel}
\usepackage{xstring}
\usepackage{axodraw2}
\usepackage[shortlabels]{enumitem}
\usepackage{siunitx}

\renewcommand{\ref}[1]{(\ref{#1})}

\newcommand{\Nf}{\ensuremath{N_{f}}}
\newcommand{\Nc}{\ensuremath{N_{c}}}
\newcommand{\Ng}{\ensuremath{N_{\gamma}}}
\DeclareMathOperator{\tr}{\mathrm{tr}}

\DeclareMathOperator{\Det}{\mathrm{Det}}
\newcommand{\ii}{\ensuremath{\mathrm{i}}}
\newcommand{\dr}{\ensuremath{\mathrm{d}}}

\newcommand{\Mz}{\ensuremath{M_0}}
\newcommand{\seff}{S_{\text{eff}}}

\DeclareMathOperator*{\argminA}{argmin} 
\providecommand{\Rcite}[1]{%
	\begingroup
	\def\tempx{0}%
	\StrCount{#1}{,}[\tempx]%
	\ifnum\tempx > 0 
	Refs.~%
	\else
	Ref.~%
	\fi
	\endgroup
	\cite{#1}%
}

\definecolor{c0}{HTML}{f36d26}
\definecolor{c1}{HTML}{178d84}
\definecolor{c2}{HTML}{bd4370}
\definecolor{c3}{HTML}{1c5b9f}
\definecolor{c4}{HTML}{569c2a}
\definecolor{c5}{HTML}{daa813}
\definecolor{c6}{HTML}{9772db}
\definecolor{c7}{HTML}{7a5d46}

\begin{document}
	\maketitle


\section{Introduction}

	In an inhomogeneous chiral phase both chiral symmetry and translational symmetry are broken by a non-zero spatially oscillating chiral condensate.
	Such a phase is a common feature in four-fermion and related Yukawa type models at non-zero quark chemical potential, e.g.\ the $1+1$-dimensional Gross-Neveu model \cite{Thies:2003kk} or the $3+1$-dimensional Nambu-Jona-Lasinio (NJL) model \cite{Nakano:2004cd,Nickel:2009wj,Sadzikowski:2000ap}.
	Interestingly, such a scenario might also be realized in QCD, since an investigation based on the functional renormalization group has found a moat regime in the QCD phase diagram \cite{Fu:2019hdw}, which might be a precursor of an inhomogeneous phase \cite{Pisarski:2021qof}. Similarly, Dyson-Schwinger calculations in QCD found an inhomogeneous phase at low temperature and large chemical potential featuring a chiral density wave within the employed truncation \cite{Muller:2013tya}.
	
	The majority of existing model calculations with focus on inhomogeneous phases are restricted to the mean-field approximation, i.e.\ neglecting bosonic quantum fluctuations.
	Recently, however, lattice Monte-Carlo simulations of $1+1$-dimensional models were successfully carried out and provided evidence that inhomogeneous regimes also exist in full quantum field theories \cite{Lenz:2020bxk,Lenz:2020cuv,Lenz:2021kzo,Nonaka:2021pwm}. 
	Extending these simulations to $3+1$-dimensional models that are closer to QCD, such as the $3+1$-dimensional NJL model, is straightforward from a technical point of view, but computationally very expensive.
	As a preparatory step it is, thus, appropriate to carry out corresponding lattice mean-field analyses, to test the lattice setup and to clarify possibly existing conceptual questions.
	In this work we report on such mean-field analyses of the $3+1$-dimensional NJL model. Since the model is not renormalizable, our main focus has been to clarify, to what extent inhomogeneous phases depend on the regularization scheme. In this way our investigation is similar to the investigation in \Rcite{Partyka:2008sv}, but with focus on lattice regularizations.


\section{\label{SEC002}The $3+1$-dimensional Nambu-Jona-Lasinio model}

	The action of the $3+1$-dimensional NJL model in the chiral limit with a quark chemical potential $\mu$ at temperature $T=1/\beta$ is given by
	\begin{align}
		\label{eq:fermi_action}
		S[\bar{\psi},\psi] = \int \dr ^3 x \int_0^\beta \dr t\, \left\{ \bar\psi \left( \slashed \partial + \gamma_0 \mu \right) \psi + G \left[\left( \bar\psi \psi\right)^2 + \left(\bar\psi \ii \gamma_5 \pmb \tau \psi\right)^2 \right]\right\}, 
	\end{align}
	where $\bar\psi$ and $\psi$ are fermion fields with $\Nf \times \Nc \times \Ng = 2 \times 3 \times 4$ components (representing the number of massless flavors, colors and spin components, respectively), $G$ is the coupling of the four-fermion interaction and $\pmb \tau$ is the vector of Pauli matrices acting in isospin space.
	
	Introducing bosonic auxiliary fields $\sigma$ and $\pmb \pi$ via a Hubbard-Stratonovich transformation and integrating over the fermion fields leads to the effective action
	\begin{align}
		\label{eq:S_eff} \seff[\sigma, \pmb \pi] = \int \dr ^3 x \int_0^\beta \dr \tau \ \frac{\sigma^2 + \pmb \pi ^2}{4G}  - \ln \Det\, D \quad , \quad Z = \int \mathcal{D}\sigma \, \mathcal{D}\pmb \pi\, e^{-\seff[\sigma,\pmb \pi]} ,
	\end{align}
	where
	\begin{equation}
		\label{EQN001} D = \slashed \partial+\gamma_0\mu  + m_0 + \sigma + \ii \gamma_5\, \pmb \tau \cdot \pmb \pi
	\end{equation}
	is the Dirac operator. The effective action is real-valued for all values of the chemical potential $\mu$ and all field configurations $(\sigma, \pmb \pi)$ (see e.g.\ \Rcite{Hands:2001aq}).
	The expectation values of the bosonic fields satisfy the Ward identities $\langle \bar \psi \psi \rangle = - 2G \langle \sigma \rangle$ and $\langle \bar \psi \ii \gamma_5 \tau_i \psi \rangle = - 2G \langle \pi_i \rangle$. One can, thus, study the purely bosonic theory (\ref{eq:S_eff}) and still obtain insights concerning the chiral condensate and chiral symmetry breaking.

	We treat this model in the mean-field approximation.
	This amounts to considering only a single field configuration $(\sigma, \pmb \pi)$, which minimizes $\seff$ globally, instead of summing over all field configurations as e.g.\ in the partition function $Z$.
	
	The $3+1$-dimensional NJL model is non-renormalizable, i.e.\ the regulator cannot be removed via renormalization. The model, thus, has two parameters: the coupling $G$ and the regulator $\Lambda$\footnote{In this work we compare several regularization schemes, where each regularization scheme defines the regulator in a different way. For simplicity we denote all regulators by $\Lambda$. For example, $\Lambda$ can represent a momentum cutoff or be proportional to the inverse lattice spacing.}.
	These parameters are tuned in such a way that the pion decay constant $f_\pi$ and the constituent quark mass $\Mz$ (in this case the chiral condensate in the vacuum) assume physically motivated values.
	By choosing $f_\pi = 88 \, \mathrm{MeV}$ and varying $\Mz$ in the range $160 \ldots 450 \, \mathrm{MeV}$ we follow a procedure outlined in \Rcite{Klevansky:1992qe,Hands:2004uv}.
	We note that decreasing $\Mz$ leads to increasing $\Lambda$.


\section{Stability analyses}

	One can explore the existence of inhomogeneous phases in a comparatively easy and cheap way via stability analyses.
	The basic idea is to consider arbitrary infinitesimal inhomogeneous perturbations for a given homogeneous field configuration $(\sigma, \pmb \pi) = (\bar{\sigma}, \bar{\pmb \pi})$ and to determine the curvature of the effective action $\seff$ with respect to these perturbations.
	If $(\bar{\sigma}, \bar{\pmb \pi})$ corresponds to the global homogeneous minimum (i.e.\ the field configuration from the set of homogeneous field configurations that minimizes $\seff$), a negative curvature indicates that there is an energetically preferred inhomogeneous condensate.
	This curvature is given by the bosonic two-point function 
	\begin{align}
		\nonumber
		\Gamma^{(2)}_\phi(q)& = \frac{1}{2G} \ +\  \begin{axopicture}(100,30)
			\Text(50,25){$p$}
			\Text(50,-25){$p+ q$}		
			\Text(17.5,11){$q$}
			\Text(82.5,11){$q$}
			\Text(17.5,-11){$\phi$}
			\Text(82.5,-11){$\phi$}
			\Text(50,0){$\psi$}
			\SetColor{c0}
			\Arc[arrow,width=1](50,0)(15,0,180)
			\Arc[arrow,width=1](50,0)(15,180,360)
			\SetColor{c1}
			\DashLine[width=2](0,0)(35,0){2}
			\DashLine[width=2](65,0)(100,0){2}
			\SetColor{black}
			\Vertex(35,0){2}
			\Vertex(65,0){2}
		\end{axopicture} = \\[8mm]
		\label{eq:twp_diagram}
		&= \frac{1}{2G}+ \int \frac{\dr^3 p}{(2 \uppi)^3} \frac{1}{\beta} \sum_n \tr \left[ S\left(p+q,\sqrt{\bar \sigma^2 + \bar {\pmb \pi}^2}\right) \, c_\phi \, S\left(p,\sqrt{\bar \sigma^2 + \bar {\pmb \pi}^2}\right) \, c_\phi \right] ,
	\end{align}
	where $\phi$ represents one of the four components of $(\sigma, \pmb \pi)$, $q = (0, \pmb q)$ is the momentum of the inhomogeneous perturbation, $p = (\omega_n,\pmb p)$ with $\omega_n$ denoting Matsubara frequencies, $S(p,\bar \sigma)$ is the fermionic propagator, $c_\sigma = \mathbb{1}$, $c_{\pi_j} = \ii \gamma_5 \tau_j$ and $\tr$ denotes the trace in isospin, color and spin space. We note that $\Gamma^{(2)}_\sigma(q) = \Gamma^{(2)}_{\pi_j}(q)$ for $(\bar \sigma, \bar{\pmb \pi}) = 0$.

	In \cref{SEC001} we present instability regions in the $\Mz$-$\mu$ plane, which are defined by having at least one direction of negative curvature, i.e.\ where $\Gamma^{(2)}_\phi(q) < 0$ for at least one $\phi$ and $q \neq 0$. Such instability regions either correspond to or are part of inhomogeneous phases. 
	For a more detailed discussion of stability analyses we refer to \Rcite{Koenigstein:2021llr,Buballa:2020nsi,Winstel:2022jkk}.


\section{\label{sec:regs}Regularization schemes}

	We carry out computations with two continuum and three lattice regularization schemes, which require different expressions for the propagator $S$ and the integration over fermionic momenta $\int \frac{\dr^3 p}{(2 \uppi)^3} \frac{1}{\beta} \sum_n$ appearing e.g.\ in \cref{eq:twp_diagram}.
	This is briefly summarized in \cref{tab:regTable}.
	
	\begin{table}
		\centering
		\begin{tabular}{c|c|c}
			regularization & $S(p,\bar \sigma)$  & momentum integration \\
			\hline
			\hline
			Pauli-Villars & $\vphantom{\Bigg|}\frac{-\ii \slashed p +\bar \sigma}{p^2 + \bar \sigma^2}$  & $\frac{1}{\beta} \sum_{n=-\infty}^{\infty} \int \dr^3 p \left(f(\omega_n,\pmb  p,  \bar \sigma) - \sum_{j=1}^{N_{\mathrm{PV}}} a_j f(\omega_n,\pmb  p, m_j \right)$  \\
			\hline
			\parbox[c]{3.0cm}{ Spatial \\momentum cutoff} & $\vphantom{\bigg|}\frac{-\ii \slashed p +\bar \sigma}{p^2 +\bar \sigma^2}$ & $\frac{1}{\beta}\sum_{n=-\infty}^{\infty} \int_0^{2\uppi} \dr \phi \, \int_0^\uppi \dr \theta \, \sin \theta \int_0^\Lambda \dr p \, p^2 f(\omega_n,\pmb  p, \bar\sigma)$ \\
			\hline
			SLAC& $\vphantom{\bigg|}\frac{-\ii \slashed p + \bar\sigma}{p^2 +\bar \sigma^2}$ & $	\frac{1}{a^4} \sum_{p\, \in\, \Gamma_{\mathrm{SLAC}}} f(\omega_n,\pmb  p, \bar\sigma) $ \\
			\hline
			\parbox[c]{3.0cm}{Hybrid with $\tilde W_\Theta(p)$ \\ Hybrid with $\tilde W_c(p)$} & $\vphantom{\bigg|}\frac{-\ii (\gamma_0 p_0 + \gamma_j \sin (p_j))+\bar \sigma}{p_0^2 + \sin^2 (p_i) +\bar \sigma^2}$ & $	\frac{1}{a^4} \sum_{p\, \in\, \Gamma_{\mathrm{Hybrid}}} f(\omega_n,\pmb  p, \bar\sigma)$  \\
			\hline
		\end{tabular}
	\caption{\label{tab:regTable}Summary of the main equations corresponding to the five regularization schemes used in this work: propagator $S$ and momentum integration of a generic function $f$ of the fermionic Matsubara frequencies $\omega_n$, the fermionic momentum $\pmb p$ and a mass $\bar \sigma$.}
	\end{table}


	\subsection{Continuum regularization schemes}
	
	\paragraph{Pauli-Villars}

	The Pauli-Villars (PV) regularization scheme removes UV-divergencies in momentum integrals by introducing additional propagators with mass $m_j = \sqrt{\bar \sigma^2 + \alpha_j \Lambda^2}$.
	This is sketched in \cref{tab:regTable}, where $N_{\mathrm{PV}} \geq 2$ is the number of PV regularization terms, $a_j$ and $\alpha_j$ are coefficients, which have to be properly tuned, and $\Lambda$ is the so-called PV mass serving as the regulator \cite{Klevansky:1992qe}.
	We choose $N_{\mathrm{PV}}=3$, $\pmb a = (1,-3,3,-1)$ and $\pmb \alpha = (0,1,2,3)$.
	The PV regularization scheme is very common in existing investigations of inhomogeneous phases in NJL-type models (see e.g.\ \Rcite{Nickel:2009wj,Heinz:2015lua}).

	\paragraph{Spatial momentum cutoff}
	
	A spatial momentum cutoff removes UV-divergencies in momentum integrals by limiting the region of integration for spatial momenta to a three dimensional sphere with radius $\Lambda$. The summation over Matsubara frequencies is not truncated. 
	This regularization scheme breaks both Euclidean rotational symmetry and translational invariance, even at $T=0$ and $\mu=0$.


\subsection{Lattice discretizations}

	\paragraph{SLAC fermions}
	SLAC fermions are common in lattice field theory and are extensively discussed in the literature (see e.g.\ \Rcite{Lenz:2020bxk} and references therein).

	\paragraph{Hybrid discretizations}
	This type of lattice discretization combines two standard approaches, the SLAC discretization in the temporal direction and the naive discretization in the three spatial directions. The latter leads to fermion doubling in the spatial dimensions, which can, however, easily be compensated in the mean field approximation by rescaling the coupling. It is, however, mandatory to modify the interaction between fermions and bosons, to suppress unphysical interactions between the doublers. This requires an appropriate weight function in momentum space, for which we consider two possibilities,
	\begin{align}
		\tilde W_\Theta(p) = \prod_{\mu=0}^{3} \Theta(\uppi/2 - |p_\mu|) \quad \text{and} \quad \tilde W_c(p) =  \prod_{\mu=0}^{3} \frac{1+\cos(p_\mu)}{2}. \label{eq:weight}
	\end{align}
	For details we refer to \Rcite{Lenz:2020bxk,Buballa:2020nsi}.

	\paragraph{Momentum integration, finite spacetime volume}	
	On a finite periodic lattice with $N_t \times N_s^3$ lattice sites and lattice spacing $a$ momentum integrations are replaced by finite sums (see \cref{tab:regTable}). The corresponding discrete momenta are
	\begin{eqnarray}
		& & \hspace{-0.7cm} \Gamma_{\mathrm{SLAC}}=\left\{p=\left(\frac{\uppi}{N_\mu a} 2n_\mu\right)\, \Big|\, n_\mu \in  \left\{\frac{N_\mu-1}{2},\frac{N_\mu-3}{2},\ldots, \frac{1-N_\mu}{2}\right\}\right\} \\
		& & \hspace{-0.7cm} \Gamma_{\mathrm{Hybrid}}=\left\{p=\left(\frac{\uppi}{N_\mu a} \left(2 n_\mu + \delta_{\mu, 0} \right)\right)\, \Big|\, n_\mu \in  \left\{\frac{N_\mu}{2}-1,\frac{N_\mu}{2}-2,\ldots, \frac{-N_\mu}{2}\right\}\right\} .
	\end{eqnarray}
	For the SLAC discretization the number of temporal lattice sites has to be even and the number of spatial lattice sites has to be odd, which corresponds to antiperiodic and periodic boundary conditions, respectively.
	The Hybrid discretization requires all $N_\mu$ to be even. 
	Momenta on the lattice are restricted by $|p_\mu| \leq \uppi/a$.
	We identify the regulator for the SLAC fermions with this upper bound, i.e.\ define $\Lambda = \uppi / a$.
	For the Hybrid discretization we define $\Lambda = \uppi / (2a)$ in order to reflect the effective doubling of the lattice spacing by the fermion doubling.
	
	\paragraph{Momentum integration, infinite spacetime volume}
	On an infinite lattice
	\begin{align}
	\label{EQN003} \int \frac{\dr^3 p}{(2 \uppi)^3} \frac{1}{\beta} \sum_n \rightarrow \int_{-\uppi/a}^{+\uppi/a} dp_0 \, \int_{-\uppi/a}^{+\uppi/a} dp_1 \, \int_{-\uppi/a}^{+\uppi/a} dp_2 \, \int_{-\uppi/a}^{+\uppi/a} dp_3 ,
	\end{align}
	i.e.\ a sum over Matsubara frequencies is replaced by an integral and the integration range is hypercubic. It is straightforward to solve such integrals numerically. Thus, we can perform calculations at $T = 0$ and infinite spatial volume.


\section{\label{SEC001}Results}

	In this section we present results from computations carried out with different regularization schemes, as discussed in \cref{sec:regs}. Unless explicitly stated otherwise, we consider $T = 0$ and set $f_\pi = 88 \, \si{MeV}$.

\begin{figure}
		\centering
		\includegraphics[clip,trim= 12mm 12mm 12mm 12mm, width=0.95\linewidth]{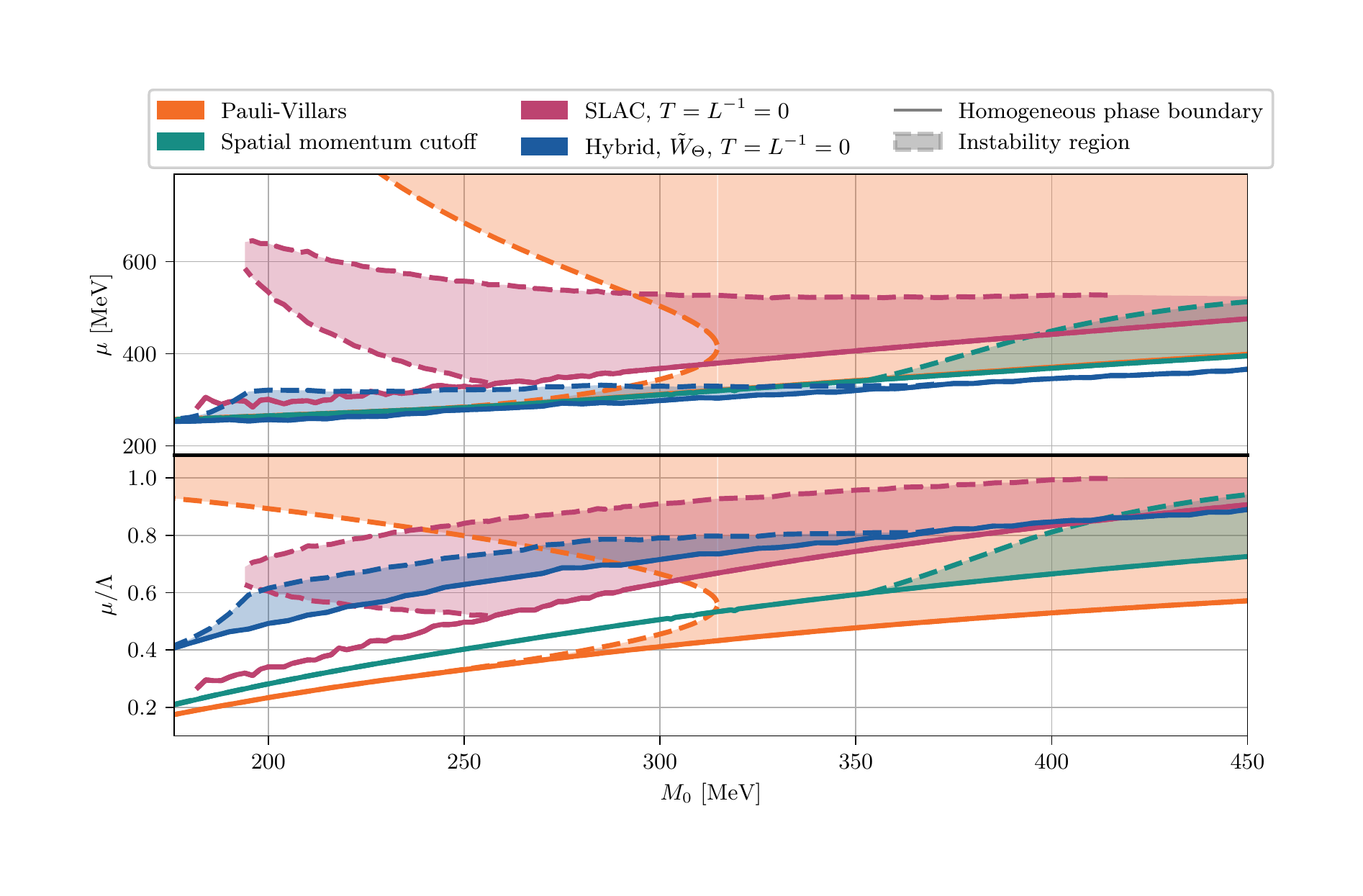}
		\caption{Homogeneous phase boundaries and instability regions in the $\Mz$-$\mu$ plane (upper plot) and in the $\Mz$-$\mu/\Lambda$ plane (lower plot) obtained with different regularization schemes for $T=0$ and $f_\pi = 88$ \, \si{MeV}. 
		For the Hybrid discretization with weight function $\tilde{W}_c$ there is only a homogeneous phase boundary, but no instability region. This homogeneous phase boundary is identical to that of the Hybrid discretization with weight function $\tilde{W}_\Theta$ and, thus, not shown in a separate color.
		Integrals (\ref{EQN003}), appearing for the lattice regularization schemes, were evaluated with statistical methods, resulting in boundaries, which exhibit small statistical fluctuations.
		}
		\label{fig:allStab}
	\end{figure}

	We obtain homogeneous phase boundaries (separating global homogeneous minima with $(\bar{\sigma}, \bar{\pmb \pi}) \neq 0$ at small chemical potential from those with $(\bar{\sigma}, \bar{\pmb \pi}) = 0$ at larger chemical potential) and instability regions in the $\Mz$-$\mu$ plane as shown in \cref{fig:allStab}. The key observations are the following:
	\begin{itemize}
	\item[\textbf{(a)}] Results obtained with different regularization schemes are in fair agreement for large values of $\Lambda$ (corresponding to values of $\Mz$, which are much smaller than the physically motivated range). This is expected, since, in general, results obtained with different regularization schemes should converge to one another for sufficiently large values of their regulators. However, for such large values of $\Lambda$, regions of instability do not exist\footnote{An exception is the disconnected instability region at very large values of $\mu$ obtained with the PV regularization -- the so-called inhomogeneous continent \cite{Buballa:2014tba}.}.

	\item[\textbf{(b)}] Four of the five regularization schemes lead to instability regions, which are drastically different. In these regions the corresponding chemical potentials are, however, of the same order or even larger than $\Lambda$ (see in particular the lower plot in \cref{fig:allStab}).
	
  \item[\textbf{(c)}] For the Hybrid discretization with weight function $\tilde{W}_c$ a region of instability does not exist.
	\end{itemize}
	From \textbf{(b)} and \textbf{(c)} we conclude that instability regions are closely related to the regulator and strongly dependent on the regularization scheme.

\begin{figure}
		\centering
		\includegraphics[width=0.47\textwidth]{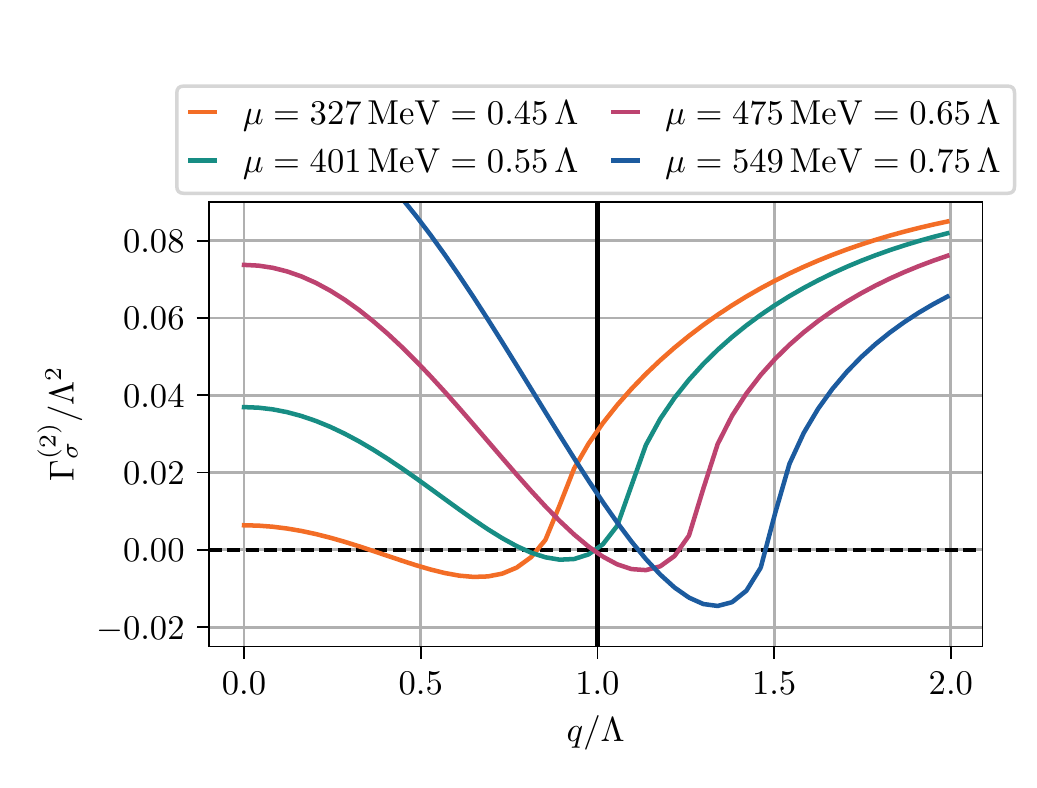}
		\includegraphics[width=0.47\textwidth]{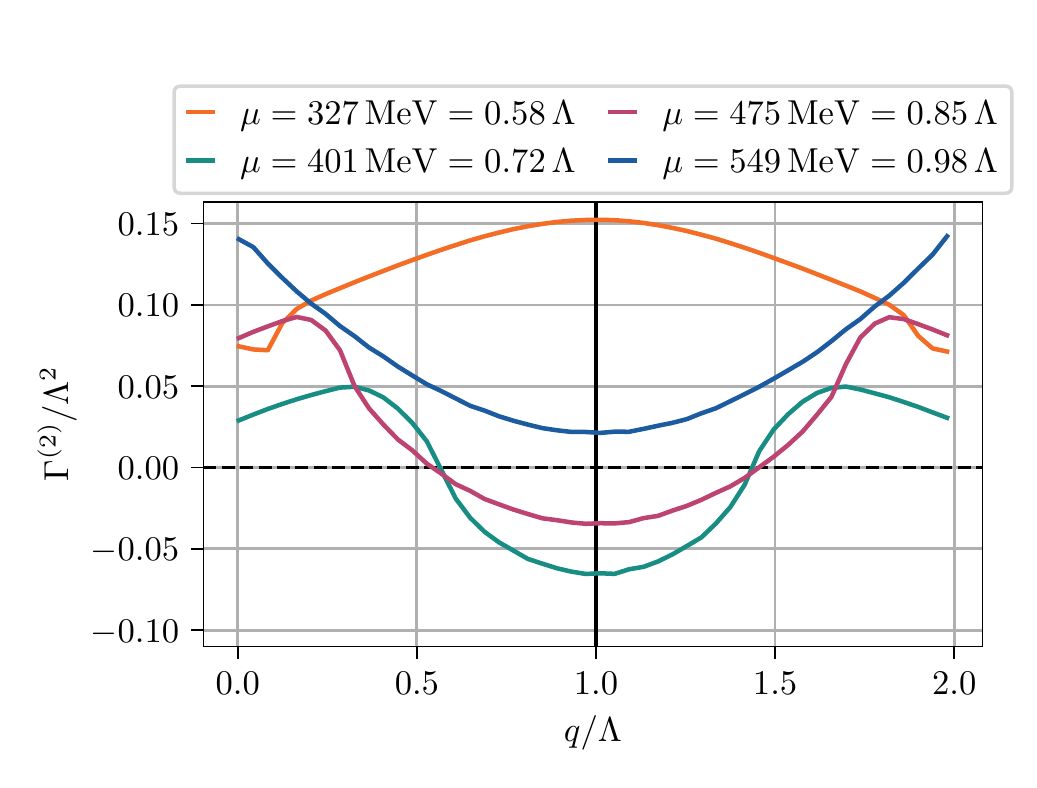}
		\caption{Two-point function $\Gamma^{(2)}_\sigma$ as function of the momentum $q$ for $T=0$, $f_\pi = 88 \, \si{MeV}$ and \\ $\Mz = 325 \, \si{MeV}$ obtained with the PV regularization (left) and the SLAC discretization (right).}
		\label{fig:PVSLACGamma2}
	\end{figure}

	To investigate this further, we define
	\begin{align}
		Q = \begin{cases}
			\argminA_{\phi,\: q > 0} \Gamma^{(2)}_\phi(q) \quad \text{if} \ \min_{\phi,\: q > 0} \Gamma^{(2)}_\phi(q) < 0 \\
			0 \quad \text{else}
		\end{cases} ,
	\end{align}
	where $\Gamma^{(2)}_\phi$ is the two-point function (\ref{eq:twp_diagram}).
	A value of $Q > 0$ indicates that the condensate is unstable with respect to inhomogeneous perturbations with momenta around $Q$ (the negative curvature of the effective action is maximized by $q=Q$). Arguments not specific to the NJL model, which are based on quark-hole pairing \cite{Buballa:2014tba}, suggest $Q \approx 2 \mu$.
	This expectation is crudely fulfilled, when using the PV regularization as illustrated by the left plot of \cref{fig:PVSLACGamma2}, where $\Gamma^{(2)}_\sigma$ is shown as function of $q$ for $\Mz = 325 \, \si{MeV}$ and several values of $\mu$. Note, however, that the momentum ranges associated with instabilities, in particular the momenta $Q$, are of the same order as the regulator $\Lambda$.
	The two-point function is symmetric in $q$ and when using a lattice discretization, $\Gamma^{(2)}_\phi$ is $2 \uppi/a$ periodic. It is thus sufficient to consider momenta $0 \leq q \leq \uppi/a$ with the lattice discretizations\footnote{The upper bound $\uppi/a$ corresponds to $\Lambda$ and $2\Lambda$ for the SLAC  and Hybrid discretization respectively.}.
	For the SLAC discretization instabilities are present, with corresponding momenta as well close to $\Lambda$ (see the right plot of \cref{fig:PVSLACGamma2}). In particular, $Q = \Lambda$.

\begin{figure}
		\centering
		\includegraphics[width=1.0 \linewidth]{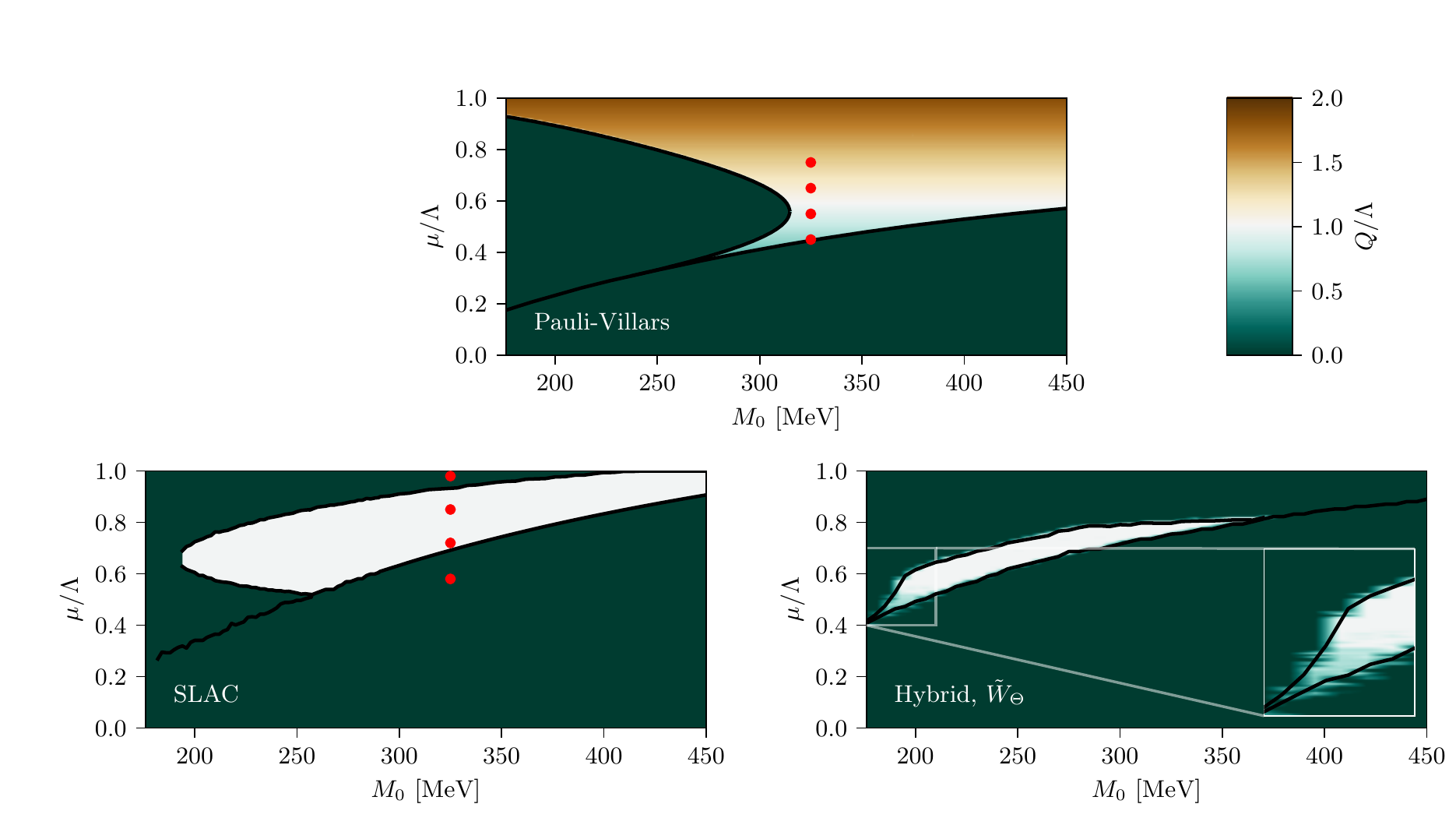}	
		\caption{$Q$ in the $\Mz$-$\mu$ plane for three different regularizations for $T = 0$ and $f_\pi = 88 \, \si{MeV}$. Red dots represent the curves from \cref{fig:PVSLACGamma2}.
		}
		\label{fig:QScan}
	\end{figure}

	Closely related is \cref{fig:QScan}, where $Q$ is shown in the $\Mz$-$\mu$ plane for the PV regularization and for the two lattice discretizations, where instability regions are found (SLAC and Hybrid with weight function $\tilde{W}_\Theta$). Both lattice discretizations lead almost exclusively to either $Q = 0$, i.e.\ no instabilities, or to $Q = \Lambda$, i.e.\ instabilities with preferred momenta close to the regulator. The PV regularization, on the other hand, leads to instability regions, where $Q \approx 2 \mu$. However, the preferred momenta are again close to the regulator or even larger, i.e.\ $Q > \Lambda$.

	Item \textbf{(b)} discussed above and the results collected in \cref{fig:PVSLACGamma2} and in \cref{fig:QScan} complement each other and provide evidence, that both $\mu$ and $Q$, two quantities closely related to the appearance of inhomogeneous phases, are of similar magnitude or larger than the regulator $\Lambda$. This raises doubts concerning the physical implication of results on inhomogeneous phases in the NJL model and explains that different regularizations lead to drastically different instability regions (see \cref{fig:allStab}).

\begin{figure}
		\centering
		\includegraphics[width=0.48\linewidth]{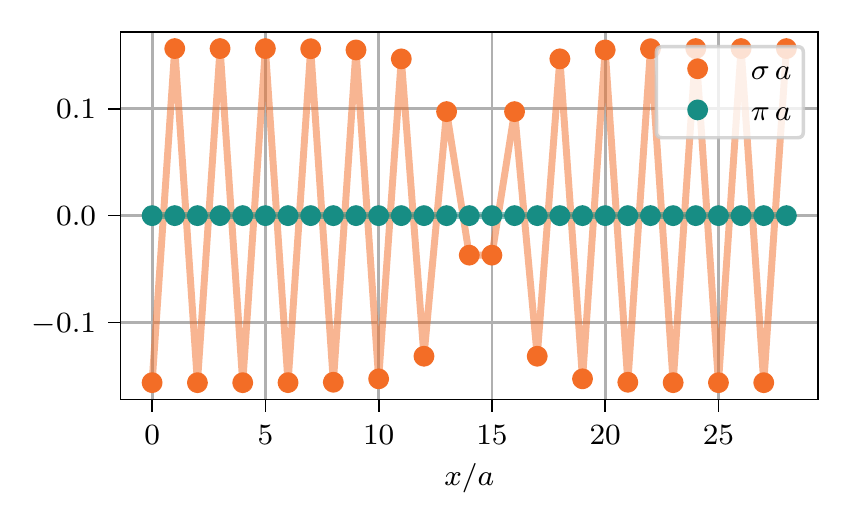}
		\includegraphics[width=0.48\linewidth]{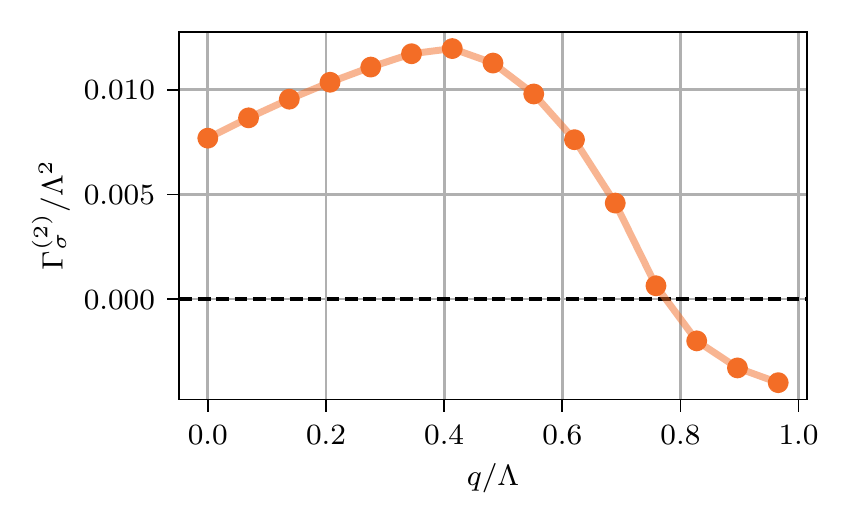}
		\caption{SLAC regularization, $T = L^{-1} \approx 7.4 \, \si{MeV}$, $f_\pi = 88 \, \si{MeV}$, $\Mz = 238 \, \si{MeV}$ and $\mu = 442 \, \si{MeV}$. \textbf{(left)}~Energetically preferred field configuration $(\sigma , \pmb \pi)$ as function of the spatial coordinate $x$ (arbitrary modulations were allowed in $x$ direction). \textbf{(right)}~Two-point function $\Gamma^{(2)}_\sigma$ as function of the momentum $q$.}
		\label{fig:slac1dgamma2}
	\end{figure}

	Lattice field theory offers the possibility to go beyond simple stability analyses by minimizing the effective action (\ref{eq:S_eff}) with respect to $\sigma$ and $\pmb \pi$. Such minimizations can be done numerically without providing specific ans\"atze for the fields. We present corresponding results for $T = L^{-1} \approx 7.4 \, \si{MeV}$ ($L$ denotes the extent of the $3$ periodic spatial dimensions), i.e.\ an $N_t \times N_s^3 = 30 \times 29^3$ lattice, $f_\pi = 88 \, \si{MeV}$, $\Mz = 238 \, \si{MeV}$ and $\mu = 442 \, \si{MeV}$ in the left plot of \cref{fig:slac1dgamma2}. Modulations were restricted to only one of the three spatial directions to limit the computational costs. The energetically preferred field configuration corresponds to $\pmb \pi = 0$, but to an oscillating $\sigma$ with wave length $2 a = 2 \pi / \Lambda$, which is the smallest realizable wave length on a lattice. This is another clear sign that the resulting inhomogeneous $\sigma$ condensate is not physically meaningful, but closely connected to the regulator. For completeness we show the corresponding two-point function $\Gamma^{(2)}_\sigma$ evaluated at the global homogeneous minimum $\bar \sigma = \bar \pi_i =0$ in the right plot of \cref{fig:slac1dgamma2}. It is qualitatively similar to the two-point functions at $T = 0$ (see \cref{fig:PVSLACGamma2}). In particular the minimum is again at $Q = \Lambda$.

	We note that the results presented in this section are in certain aspects similar to results obtained in the $2+1$-dimensional Gross-Neveu model \cite{Buballa:2020nsi, Winstel:2022jkk, Narayanan:2020uqt}. In these references unphysical instability regions and inhomogeneous phases were obtained with the PV regularization and the Hybrid discretization with weight function $\tilde{W}_\Theta$. It was also shown that these instability regions and inhomogeneous phases are only present at finite values of the regulators and disappear after renormalization.

	To summarize, the results presented in this section indicate that existing results on inhomogeneous phases in the $3+1$-dimensional NJL model should be taken with caution, in particular, when only a single regularization scheme was used. 
	Instability regions seem to be closely related to the regulator, which could be an indication that inhomogeneous phases are rather generated by the regularization and not by the NJL Lagrangian and the symmetries it shares with QCD.	Thus, it seems questionable, whether such results have any implication or could provide hints on the existence of inhomogeneous phases in QCD.


\section*{Acknowledgments}

	We acknowledge useful discussions with J.~Braun, M.~Buballa, Ph.~de Forcrand, A.~Koenigstein, L.~Kurth, J.~Lenz, M.~Mandl, G.~D.~Moore, J.~Pawlowski, O.~Philipsen, F.~Rennecke and A.~Wipf.	
	This work was supported by the Deutsche Forschungsgemeinschaft (DFG, German Research Foundation) – project number 315477589 – TRR 211.
	M.~Wagner acknowledges support by the Heisenberg Programme of the Deutsche Forschungsgemeinschaft (DFG, German Research Foundation) -- project number 399217702. 
	M.~Winstel acknowledges support by the GSI Forschungs- und Entwicklungsvereinbarungen (GSI F\&E) and by the Giersch Foundation.
	Calculations on the GOETHE-HLR and on the on the FUCHS-CSC high-performance computers of the Frankfurt University were conducted for this research.
	We would like to thank HPC-Hessen, funded by the State Ministry of Higher Education, Research and the Arts, for programming advice.


\bibliographystyle{JHEP}
\bibliography{main}

\providecommand{\href}[2]{#2}\begingroup\raggedright\begin{thebibliography}{10}

\bibitem{Thies:2003kk}
M.~Thies and K.~Urlichs, \emph{Revised phase diagram of the {{Gross-Neveu}}
  model}, \href{https://doi.org/10.1103/PhysRevD.67.125015}{\emph{Physical
  Review} {\bfseries D67} (2003) 125015}
  [\href{https://arxiv.org/abs/hep-th/0302092}{{\ttfamily hep-th/0302092}}].

\bibitem{Nakano:2004cd}
E.~Nakano and T.~Tatsumi, \emph{Chiral symmetry and density wave in quark
  matter}, \href{https://doi.org/10.1103/PhysRevD.71.114006}{\emph{Physical
  Review D} {\bfseries 71} (2005) 114006}
  [\href{https://arxiv.org/abs/hep-ph/0411350}{{\ttfamily hep-ph/0411350}}].

\bibitem{Nickel:2009wj}
D.~Nickel, \emph{Inhomogeneous phases in the {{Nambu-Jona-Lasino}} and
  quark-meson model},
  \href{https://doi.org/10.1103/PhysRevD.80.074025}{\emph{Physical Review}
  {\bfseries D80} (2009) 074025}
  [\href{https://arxiv.org/abs/0906.5295}{{\ttfamily 0906.5295}}].

\bibitem{Sadzikowski:2000ap}
M.~Sadzikowski and W.~Broniowski, \emph{Non-uniform chiral phase in effective
  chiral quark models},
  \href{https://doi.org/10.1016/S0370-2693(00)00830-3}{\emph{Physics Letters B}
  {\bfseries 488} (2000) 63}
  [\href{https://arxiv.org/abs/hep-ph/0003282}{{\ttfamily hep-ph/0003282}}].

\bibitem{Fu:2019hdw}
W.-j.~Fu, J.M.~Pawlowski and F.~Rennecke, \emph{The {{QCD}} phase structure at
  finite temperature and density},
  \href{https://doi.org/10.1103/PhysRevD.101.054032}{\emph{Physical Review D}
  {\bfseries 101} (2020) 054032}
  [\href{https://arxiv.org/abs/1909.02991}{{\ttfamily 1909.02991}}].

\bibitem{Pisarski:2021qof}
R.D.~Pisarski and F.~Rennecke, \emph{Signatures of {{Moat Regimes}} in
  {{Heavy-Ion Collisions}}},
  \href{https://doi.org/10.1103/PhysRevLett.127.152302}{\emph{Physical Review
  Letters} {\bfseries 127} (2021) 152302}
  [\href{https://arxiv.org/abs/2103.06890}{{\ttfamily 2103.06890}}].

\bibitem{Muller:2013tya}
D.~M{\"u}ller, M.~Buballa and J.~Wambach, \emph{Dyson-{{Schwinger}} study of
  chiral density waves in {{QCD}}},
  \href{https://doi.org/10.1016/j.physletb.2013.10.050}{\emph{Physics Letters
  B} {\bfseries 727} (2013) 240}
  [\href{https://arxiv.org/abs/1308.4303}{{\ttfamily 1308.4303}}].

\bibitem{Lenz:2020bxk}
J.~Lenz, L.~Pannullo, M.~Wagner, B.~Wellegehausen and A.~Wipf,
  \emph{Inhomogeneous phases in the {{Gross-Neveu}} model in \$1+1\$ dimensions
  at finite number of flavors},
  \href{https://doi.org/10.1103/PhysRevD.101.094512}{\emph{Physical Review D}
  {\bfseries 101} (2020) 094512}.

\bibitem{Lenz:2020cuv}
J.J.~Lenz, L.~Pannullo, M.~Wagner, B.~Wellegehausen and A.~Wipf, \emph{Baryons
  in the {{Gross-Neveu}} model in 1+1 dimensions at finite number of flavors},
  \href{https://doi.org/10.1103/PhysRevD.102.114501}{\emph{Physical Review D}
  {\bfseries 102} (2020) 114501}
  [\href{https://arxiv.org/abs/2007.08382}{{\ttfamily 2007.08382}}].

\bibitem{Lenz:2021kzo}
J.J.~Lenz, M.~Mandl and A.~Wipf, \emph{Inhomogeneities in the \$2\$-{{Flavor
  Chiral Gross-Neveu Model}}}, {\emph{arXiv:2109.05525 [cond-mat,
  physics:hep-lat, physics:hep-th]} (2021) }
  [\href{https://arxiv.org/abs/2109.05525}{{\ttfamily 2109.05525}}].

\bibitem{Nonaka:2021pwm}
C.~Nonaka and K.~Horie, \emph{Inhomogeneous phases in the chiral
  {{Gross-Neveu}} model on the lattice},
  \href{https://doi.org/10.22323/1.396.0150}{\emph{PoS} {\bfseries LATTICE2021}
  (2022) 150}.

\bibitem{Partyka:2008sv}
T.L.~Partyka and M.~Sadzikowski, \emph{Phase diagram of the non-uniform chiral
  condensate in different regularization schemes at {{T}}=0},
  \href{https://doi.org/10.1088/0954-3899/36/2/025004}{\emph{Journal of Physics
  G: Nuclear and Particle Physics} {\bfseries 36} (2009) 025004}
  [\href{https://arxiv.org/abs/0811.4616}{{\ttfamily 0811.4616}}].

\bibitem{Hands:2001aq}
S.~Hands, B.~Lucini and S.~Morrison, \emph{Numerical {{Portrait}} of a
  {{Relativistic Thin Film BCS Superfluid}}},
  \href{https://doi.org/10.1103/PhysRevD.65.036004}{\emph{Physical Review D}
  {\bfseries 65} (2002) 036004}
  [\href{https://arxiv.org/abs/hep-lat/0109001}{{\ttfamily hep-lat/0109001}}].

\bibitem{Klevansky:1992qe}
S.P.~Klevansky, \emph{The {{Nambu-Jona-Lasinio}} model of quantum
  chromodynamics},
  \href{https://doi.org/10.1103/RevModPhys.64.649}{\emph{Reviews of Modern
  Physics} {\bfseries 64} (1992) 649}.

\bibitem{Hands:2004uv}
S.~Hands and D.N.~Walters, \emph{Numerical {{Portrait}} of a {{Relativistic BCS
  Gapped Superfluid}}},
  \href{https://doi.org/10.1103/PhysRevD.69.076011}{\emph{Physical Review D}
  {\bfseries 69} (2004) 076011}
  [\href{https://arxiv.org/abs/hep-lat/0401018}{{\ttfamily hep-lat/0401018}}].

\bibitem{Koenigstein:2021llr}
A.~Koenigstein, L.~Pannullo, S.~Rechenberger, M.~Winstel and M.J.~Steil,
  \emph{Detecting inhomogeneous chiral condensation from the bosonic two-point
  function in the \$(1 + 1)\$-dimensional {{Gross-Neveu}} model in the
  mean-field approximation},
  \href{https://doi.org/10.1088/1751-8121/ac820a}{\emph{Journal of Physics A:
  Mathematical and Theoretical} {\bfseries 55} (2022) 375402}
  [\href{https://arxiv.org/abs/2112.07024}{{\ttfamily 2112.07024}}].

\bibitem{Buballa:2020nsi}
M.~Buballa, L.~Kurth, M.~Wagner and M.~Winstel, \emph{Regulator dependence of
  inhomogeneous phases in the 2+1-dimensional {{Gross-Neveu}} model},
  \href{https://doi.org/10.1103/PhysRevD.103.034503}{\emph{Physical Review D}
  {\bfseries 103} (2021) 034503}
  [\href{https://arxiv.org/abs/2012.09588}{{\ttfamily 2012.09588}}].

\bibitem{Winstel:2022jkk}
M.~Winstel and L.~Pannullo, \emph{Stability of homogeneous chiral phases
  against inhomogeneous perturbations in 2+1 dimensions},  Nov., 2022.
\newblock 10.48550/arXiv.2211.04414.

\bibitem{Heinz:2015lua}
A.~Heinz, F.~Giacosa, M.~Wagner and D.H.~Rischke, \emph{Inhomogeneous
  condensation in effective models for {{QCD}} using the finite-mode approach},
  \href{https://doi.org/10.1103/PhysRevD.93.014007}{\emph{Physical Review}
  {\bfseries D93} (2016) 014007}
  [\href{https://arxiv.org/abs/1508.06057}{{\ttfamily 1508.06057}}].

\bibitem{Buballa:2014tba}
M.~Buballa and S.~Carignano, \emph{Inhomogeneous chiral condensates},
  \href{https://doi.org/10.1016/j.ppnp.2014.11.001}{\emph{Progress in Particle
  and Nuclear Physics} {\bfseries 81} (2015) 39}
  [\href{https://arxiv.org/abs/1406.1367}{{\ttfamily 1406.1367}}].

\bibitem{Narayanan:2020uqt}
R.~Narayanan, \emph{Phase diagram of the large \${{N}}\$ {{Gross-Neveu}} model
  in a finite periodic box},
  \href{https://doi.org/10.1103/PhysRevD.101.096001}{\emph{Phys. Rev. D}
  {\bfseries 101} (2020) 096001}.

\end{thebibliography}\endgroup
\end{document}